\documentclass[apj]{emulateapj}
\usepackage{graphicx}
\usepackage{natbib}
\usepackage{amssymb}

\begin{document}

\title{Testing universal relations of neutron stars with a nonlinear
matter-gravity coupling theory}

\author{Y.-H. Sham\footnote{Email address: yhsham@phy.cuhk.edu.hk},
 L.-M. Lin\footnote{Email address: lmlin@phy.cuhk.edu.hk},
and P.~T. Leung\footnote{Email address: ptleung@phy.cuhk.edu.hk} }

\affil{Department of Physics and Institute of Theoretical Physics, 
The Chinese University of Hong Kong, Shatin, Hong Kong SAR, China }

\date{\today}

\begin{abstract}
Due to our ignorance of the equation of state (EOS) beyond nuclear density, 
there is still no unique theoretical model for neutron stars (NSs). 
It is therefore surprising that universal EOS-independent relations connecting 
different physical quantities of NSs can exist. 
Lau et al. found that the frequency of the 
$f$-mode oscillation, the mass, and the moment of inertia are connected by 
universal relations.
More recently, Yagi and Yunes discovered the 
I-Love-Q universal relations among the mass, the moment of inertia, the 
Love number, and the quadrupole moment. 
In this paper, we study these universal relations in the Eddington-inspired 
Born-Infeld (EiBI) gravity. This theory differs from general relativity (GR) 
significantly only at high densities due to the nonlinear coupling 
between matter and gravity. 
It thus provides us an ideal case to test how robust the universal relations 
of NSs are with respect to the change of the gravity theory. 
Thanks to the apparent EOS formulation of EiBI gravity developed recently
by Delsate and Steinhoff, we are able
to study the universal relations in EiBI gravity using the same techniques
as those in GR. 
We find that the universal relations in EiBI gravity are essentially
the same as those in GR. 
Our work shows that, within the currently viable coupling constant, 
there exists at least one modified gravity theory that is indistinguishable 
from GR in view of the unexpected universal relations.

\end{abstract}


\maketitle


\section{Introduction}
\label{sec:intro}

Neutron stars (NSs) are compact objects that have densities several 
times the normal nuclear density in their cores.  
Due to their internal high-density environment, NSs have long been regarded as 
the most natural cosmic laboratories for studying dense nuclear matter, which 
is not well understood and poorly constrained by experiments performed on 
Earth. 
It is hoped that inferring global properties of NSs from observations may 
allow one to select the {\it correct} equation of state (EOS) among many 
theoretical possibilities. However, achieving this goal in practice is  
non-trivial. The reason is that it is in general not possible to extract all 
global properties of a NS simultaneously from observations. 
It is thus of great interest to search for empirical relationships, which 
are in general EOS dependent (e.g., the mass-radius relation of NSs), 
connecting only a few physical quantities of a NS so that one might hope 
to put constraints on the EOS by measuring these quantities. 
On the other hand, from a fundamental physics point of view, it would even be 
more interesting if the relationships are insensitive to the NS EOS models, 
since one could then use the relations to test the underlying gravitational 
theory despite our ignorance of dense nuclear matter
\citep{Yagi:2013long,Yagi:2013}. 

In the past decade, several empirical relations connecting different 
physical parameters of NSs have been proposed. \citet{Bejger:2002p8392} and 
\citet{Lattimer:2005p7082} discovered relationships relating the scaled 
moment of inertia $I/M R^2$ and compactness $M/R$, where 
$I$ is the moment of inertia, $M$ is the mass, and $R$ is the radius.
Making use of the discovered
relation, \citet{Lattimer:2005p7082} suggested that the moment
of inertia of star A in the double pulsar system J0737-3039 
\citep{Burgay:2003p531,Lyne:2004p1153} could be determined to about 
10\% accuracy.

On the other hand, pulsating NSs are expected to be promising 
sources of gravitational waves. It is expected that studying the 
gravitational-wave signals emitted by oscillating NSs can yield 
useful information about the internal structure of the stars. 
Several universal behaviors of the quadrupolar $f$-mode have been established 
\citep{Andersson:1996p20,Andersson:1998p1059,Benhar1999:p797,
Benhar:2004:p124015,Tsui:2005p1029,Lau:2010p1234}. 
In particular, \citet{Lau:2010p1234} found a pair of nearly EOS-independent
relations to connect the frequency and damping rate of the $f$-mode to the 
global properties $M$ and $I$ of the stars (see Section~\ref{sec:fmode}). 
It has been furthermore shown \citep{Lau:2010p1234} that the values of 
$M$, $R$, and $I$ of a NS can be inferred accurately from the $f$-mode 
gravitational wave signals.

More recently, \citet{Urbanec:13p1903} discovered a universal relation 
between $QM/J^2$ and $M/R$, where $Q$ is the spin-induced quadrupole moment 
and $J$ is the angular momentum. 
\citet{Yagi:2013long,Yagi:2013} discovered universal relations 
relating $I$, $Q$, the tidal Love number $\lambda_{\rm tid}$, and the
rotational Love number $\lambda_{\rm rot}$.  
These newly discovered I-Love-Q relations will be directly 
relevant to the understanding of the gravitational wave signals emitted 
during the late stages of NS-NS binary mergers (see Section~\ref{sec:ILoveQ}).
Finally, \citet{Baubock:1306.0569} found universal relations among 
$I$, $J$, $Q$, $M/R$, and the ellipticity of the stellar surface.

While \citet{Yagi:2013long,Yagi:2013} have also studied the I-Love relation in 
an alternative theory of gravity (see Section~\ref{sec:ILoveQ}), most of 
the universal relations of NSs discussed above are based on the assumption 
that gravity is described by the theory of general relativity (GR). 
But how well do we understand gravity? 
So far the most successful theory of gravity is GR, and it has been well 
tested in weak-field situations 
\citep[for a review see][]{Will:2006}.
However, whether gravity behaves as GR predicts in strong-field situations,
such as NS-NS binary mergers, is still an open question. 
It is hopeful that testing GR in the strong-field limit will soon become 
possible in the coming decade through gravitational wave observations by 
ground based detectors such as Advanced LIGO, Advanced VIRGO, and KAGRA 
\citep{Will:1993book,Will:2006,Gair:2013,Yagi:2013review,Yunes:2013review,
Mirshekari:2013thesis}.
On the other hand, we also know that GR is not complete because 
of its prediction of singularities in the Big Bang and those inside black 
holes. While it is generally believed that quantum gravity is needed to 
resolve these problems, it is still interesting to search for alternative 
theories of gravity that could avoid the singularity problems within the 
classical level. 

In recent years, a new theory of gravity called Eddington-inspired Born-Infeld 
(EiBI) gravity proposed by \citet{Banados10} has been gaining attention
\citep[see also][]{Deser98,Vollick04}. 
EiBI gravity is appealing because it reduces to GR in vacuum and can avoid 
the Big Bang singularity \citep{Banados10}. 
The deviation between EiBI gravity and GR becomes significant only at high 
densities. 
The implications of EiBI gravity in cosmological 
\citep{Banados10,Scargill:2012p103533, Avelino:2012p041501, 
Escamilla:2012p087302, Liu:2012p124053, Cho:2012p084018, Cho:2013p071301, 
Harko:1305.0820, Bouhmadi:1302.5013}
and astrophysical \citep{Pani:2011p031101, Pani:2012p084020, Pani:2012p251102, 
Sham:2012, Sham:2013, Harko:2013p044032}
contexts have been widely investigated. 
Unlike GR, where gravity couples to matter linearly in the sense that 
the Einstein tensor $G_{\mu\nu}$ is proportional to the stress-energy tensor 
$T_{\mu\nu}$ in the Einstein field equations, EiBI gravity introduces 
nonlinear coupling between matter and gravity (see Section~\ref{sec:EiBI}). 
It is in fact the nonlinear matter-gravity coupling in this theory that is
responsible for avoiding some of the singularities that plague GR 
\citep{Delsate:2012}. 
However, it has also been shown recently that the same nonlinear coupling 
leads to some pathologies, such as surface singularities 
\citep{Pani:2012p251102} and anomalies associated with phase 
transitions \citep{Sham:2013}, for compact stars in EiBI gravity.   

While the works of \citet{Pani:2012p251102} and \citet{Sham:2013} cast doubt 
on its viability, EiBI gravity certainly stands as an interesting example of a 
more general class of nonlinear matter-gravity coupling theories, due to its 
equivalence to GR in vacuum and its ability to avoid singularities. 
As discussed above, the theory differs from GR significantly only at high 
densities. Since NSs are the most dense stellar objects in the universe, 
EiBI gravity thus provides us with an ideal test case to investigate how the 
universal relations of NSs would change if the underlying gravitational 
theory is such that matter and gravity are nonlinearly coupled together. 
In this paper, we study the $f$-mode universality relations 
and the I-Love-Q relations in EiBI gravity.  
Thanks to the apparent-EOS formulation of EiBI gravity developed by 
\citet{Delsate:2012}, 
we are able to study the universal relations in 
EiBI gravity using the same techniques as those in GR because in this 
formulation EiBI gravity coupled to a perfect fluid is equivalent to GR with 
the matter field described by an effective EOS 
(see Section~\ref{sec:EiBI_apparentEOS}). 
We find that these universal relations are indeed very robust and 
independent of whether gravity is described by GR or EiBI theory, as long 
as the coupling constant in EiBI gravity is within the range that is 
already constrained astrophysically. As we shall discuss, the results are not 
too surprising since the apparent EOS, within the range constrained 
astrophysically, does not differ significantly from nuclear EOS 
models.

The plan of the paper is as follows. We first briefly summarize EiBI gravity 
in Section~\ref{sec:EiBI}. Section~\ref{sec:EiBI_apparentEOS} discusses
the apparent EOS formulation of EiBI gravity.  
In Sections~\ref{sec:fmode} and \ref{sec:ILoveQ}, 
we review the universal behaviors of $f$-mode oscillations and the 
I-Love-Q relations for NSs discovered in GR, respectively. 
We then present our numerical results to show that these relations also 
work well in EiBI gravity. Finally, our conclusions are summarized in 
Section~\ref{sec:conclude}. 
Unless otherwise noted, we use geometric units where $G=c=1$.

\section{Eddington-Inspired Born-Infeld Gravity}
\label{sec:EiBI}

The EiBI theory is based on a Palatini formulation of the 
action \citep{Banados10} 
\begin{eqnarray}
S &=& {1 \over 16 \pi} {2 \over \kappa} \int d^4x \left( \sqrt{
\left| g_{\mu\nu} + \kappa R_{\mu\nu} \right| } - \lambda \sqrt{-
g} \right) \nonumber  \\
\cr
&& + S_M \left[ g, \Psi_M \right] ,
\label{eq:EiBI_action}
\end{eqnarray} 
where $R_{\mu\nu}$ is the symmetric part of the Ricci tensor constructed
solely by the connection $\Gamma^\alpha_{\beta\gamma}$, 
$S_M \left[ g, \Psi_M \right]$ is the matter action, and 
$|f_{\mu\nu}| \equiv f$ denotes the determinant of a tensor field 
$f_{\mu\nu}$. 
The parameters $\kappa$ and $\lambda$ are related to the cosmological constant 
$\Lambda$ by $\Lambda = (\lambda - 1)/\kappa$. 
In the limit $\kappa \rightarrow 0$, it can 
be shown that the action (Equation~(\ref{eq:EiBI_action})) reduces to the 
Einstein-Hilbert action for GR. 
We shall set $\lambda=1$ hereafter and consider $\kappa$ as the
only parameter of the theory. 
The current tightest constraint on $\kappa$ \citep{Avelino:2012p104053}
is set by the existence of NSs and is given by $8\pi\kappa\epsilon_0 < 0.1$, 
where $\epsilon_0 = 10^{15} {\ \rm g\ cm}^{-3}$.

Varying the action (Equation~(\ref{eq:EiBI_action})) with respect to the metric
$g_{\mu\nu}$ and $\Gamma^\alpha_{\beta\gamma}$ separately yields 
\begin{eqnarray}
q_{\mu\nu} &=& g_{\mu\nu} + \kappa R_{\mu\nu} , \label{eq:field1} \\
\cr
\sqrt{-q} q^{\mu\nu} &=&  \sqrt{-g} g^{\mu\nu} - 
8\pi \kappa \sqrt{-g} T^{\mu\nu} , \label{eq:field2} 
\end{eqnarray}
where $q_{\mu\nu}$ is an auxiliary metric compatible with the connection:
\begin{equation}
\Gamma^\alpha_{\beta\gamma} 
= {1 \over 2} q^{\alpha\sigma} \left( \partial_{\gamma} q_{\sigma\beta}
+\partial_{\beta} q_{\sigma\gamma} - \partial_{\sigma} q_{\beta\gamma} 
\right) . 
\end{equation}
Since the matter action $S_M$ is assumed to depend only on the metric
$g_{\mu\nu}$ and the matter fields, but not on the connection
$\Gamma^\alpha_{\beta\gamma}$, the stress-energy tensor $T^{\mu\nu}$ still 
satisfies the same conservation equations as in GR 
\begin{equation}
\nabla_\mu T^{\mu\nu} = 0 , 
\end{equation}
where the covariant derivative refers to the physical metric $g_{\mu\nu}$. 
It should be noted that $q_{\mu\nu}$ is identical to the physical 
metric $g_{\mu\nu}$ when $T^{\mu\nu}=0$. 
In fact, it can be shown that the action (Equation~(\ref{eq:EiBI_action}))
reduces to the Einstein-Hilbert action when the matter action $S_M$ vanishes 
\citep{Banados10}. Hence, EiBI gravity is identical to GR in vacuum.

\section{Apparent EOS Formulation of EiBI gravity}
\label{sec:EiBI_apparentEOS}

Solving the field equations (Equations~(\ref{eq:field1}) and 
(\ref{eq:field2})) can be more difficult than in GR. 
The reason is that the auxiliary metric $q_{\mu\nu}$ involves both 
$g_{\mu\nu}$ and $T_{\mu\nu}$ and thus the Ricci tensor $R_{\mu\nu}$ 
contains second derivatives of the matter field\footnote{Indeed, the 
dependence of $R_{\mu\nu}$ on derivatives of the matter field is the cause 
of the pathologies associated with compact stars in EiBI gravity found by
\citet{Pani:2012p251102} and \citet{Sham:2013}. }.
However, \citet{Delsate:2012} have recently shown that the field equations
(Equations~(\ref{eq:field1}) and (\ref{eq:field2})) can be written in a form 
that resembles the Einstein equations in GR:
\begin{eqnarray}
 G^\mu_\nu &\equiv& R^\mu_\nu - \frac{1}{2} \delta^\mu_\nu R \nonumber\\
&=&8\pi\tau T^\mu_\nu - \frac{1-\tau}{\kappa}\delta^\mu_\nu 
- 4\pi\tau T\delta^\mu_\nu \equiv 8\pi\tilde{T}^\mu_\nu ,
\label{eq:modified_Einstein}
\end{eqnarray}
where $R=R^\mu_\mu$ and $T=T^\mu_\mu$. 
The Einstein tensor $G^\mu_\nu$ is defined by the auxiliary metric
$q_{\mu\nu}$ and $\tilde{T}^\mu_\nu$ is defined as the apparent stress-energy 
tensor. The scalar $\tau$ is defined by $\tau = \sqrt{g/ q}$ and can be 
expressed as  
\begin{equation}
\tau = \left[ {\rm det} ( \delta^\mu_\nu - 8 \pi \kappa T^\mu_\nu )
\right]^{-1/2} . 
\end{equation}
Note that Equation~(\ref{eq:modified_Einstein}) in general still depends on 
the physical metric $g_{\mu\nu}$ through the standard stress-energy tensor
$T^\mu_\nu$. However, for the case of a perfect fluid  
\begin{equation}
T^{\mu\nu} = (\epsilon + P ) u^\mu u^\nu + P g^{\mu\nu} , 
\end{equation}
where $\epsilon$ is the energy density, $P$ is the pressure, and 
$u^\mu$ is the four velocity of the fluid, 
Equation~(\ref{eq:modified_Einstein}) can be made to decouple from 
$g_{\mu\nu}$ completely. 
In this special case, $\tau$ can be computed in a frame
comoving with the fluid and is given by   
\begin{equation} 
\tau = \left[ (1+8\pi\kappa\epsilon)(1-8\pi\kappa P)^3 \right]^{-1/2} .
\end{equation}
The apparent stress-energy tensor can be written as 
\begin{equation}
\tilde{T}^\mu_\nu = (\tilde{\epsilon} + \tilde{P}) v^\mu v_\nu 
+ \tilde{P} \delta^\mu_\nu,
\end{equation}
with the apparent energy density $\tilde{\epsilon}$ and pressure 
$\tilde{P}$ defined by 
\begin{equation}
\tilde{\epsilon} = \tau \epsilon - \bar{P}  ,
\label{eq:app_epsilon}
\end{equation}
\begin{equation}
\tilde{P} = \tau P + \bar{P}, 
\label{eq:app_pressure}
\end{equation}
where $\bar{P} \equiv (\tau-1)/8\pi\kappa - \tau(3P -\epsilon)/2$. The 
apparent four velocity $v^\mu$ satisfies the conditions 
\begin{equation}
v^\mu v^\nu q_{\mu\nu} = -1 \ \ ; \ \ v^\mu v_\nu = u^\mu u_\nu . 
\end{equation}
Note that the indices of $v^\mu$ and $u^\mu$ are lowered by $q_{\mu\nu}$ 
and $g_{\mu\nu}$, respectively. 

The advantage of reformulating EiBI gravity in the form of 
Equation~(\ref{eq:modified_Einstein}) has now become clear. 
For a given coupling parameter $\kappa$ and a given physical EOS 
$P(\epsilon)$, one can solve a problem in EiBI gravity by solving the 
same Einstein equations (Equation~\ref{eq:modified_Einstein})) as in GR, 
but with 
$q_{\mu\nu}$ as the fundamental metric and an apparent EOS 
$\tilde{P}(\tilde{\epsilon})$ given by Equations (\ref{eq:app_epsilon}) 
and (\ref{eq:app_pressure}). This implies that many theoretical ideas and 
numerical codes developed in GR can readily be transferred to EiBI gravity. 
For instance, instead of solving the field equations 
(Equations~(\ref{eq:field1}) and (\ref{eq:field2})) 
for the structure of compact stars, as was originally done
\citep{Pani:2011p031101,Pani:2012p084020,Sham:2012,Sham:2013,
Harko:2013p044032}, one can simply solve the well-known 
Tolman-Oppenheimer-Volkov (TOV) equations in GR with an apparent EOS. 

The advantage of this reformulation of EiBI gravity becomes even more 
transparent when one considers dynamical problems in EiBI gravity, where
the algebraic complexity in the analysis grows quite rapidly with the 
problem size. 
For example, it is non-trivial to calculate non-radial oscillation 
modes of NSs even in GR. Formulating the same calculation in EiBI gravity 
starting with the field equations (Equations~(\ref{eq:field1}) and 
(\ref{eq:field2})) would be a more tedious task than in GR. 
However, thanks to the apparent EOS approach, we are able 
to make use of the numerical codes we previously developed in GR 
\citep{Lau:2010p1234} to study the $f$-mode of NSs in EiBI gravity. 
As a side remark, we have indeed verified that the frequencies of the 
radial oscillation modes of NSs in EiBI gravity we obtained
previously \citep{Sham:2012}, by solving Equations~(\ref{eq:field1}) and 
(\ref{eq:field2}), agree with the results (within numerical accuracy)
obtained by solving the corresponding eigenvalue equation in GR 
\citep[see, e.g.,][]{Kokkotas:2001p565} with apparent EOSs. 

\begin{figure}
  \centering
  \includegraphics[width=7.5cm]{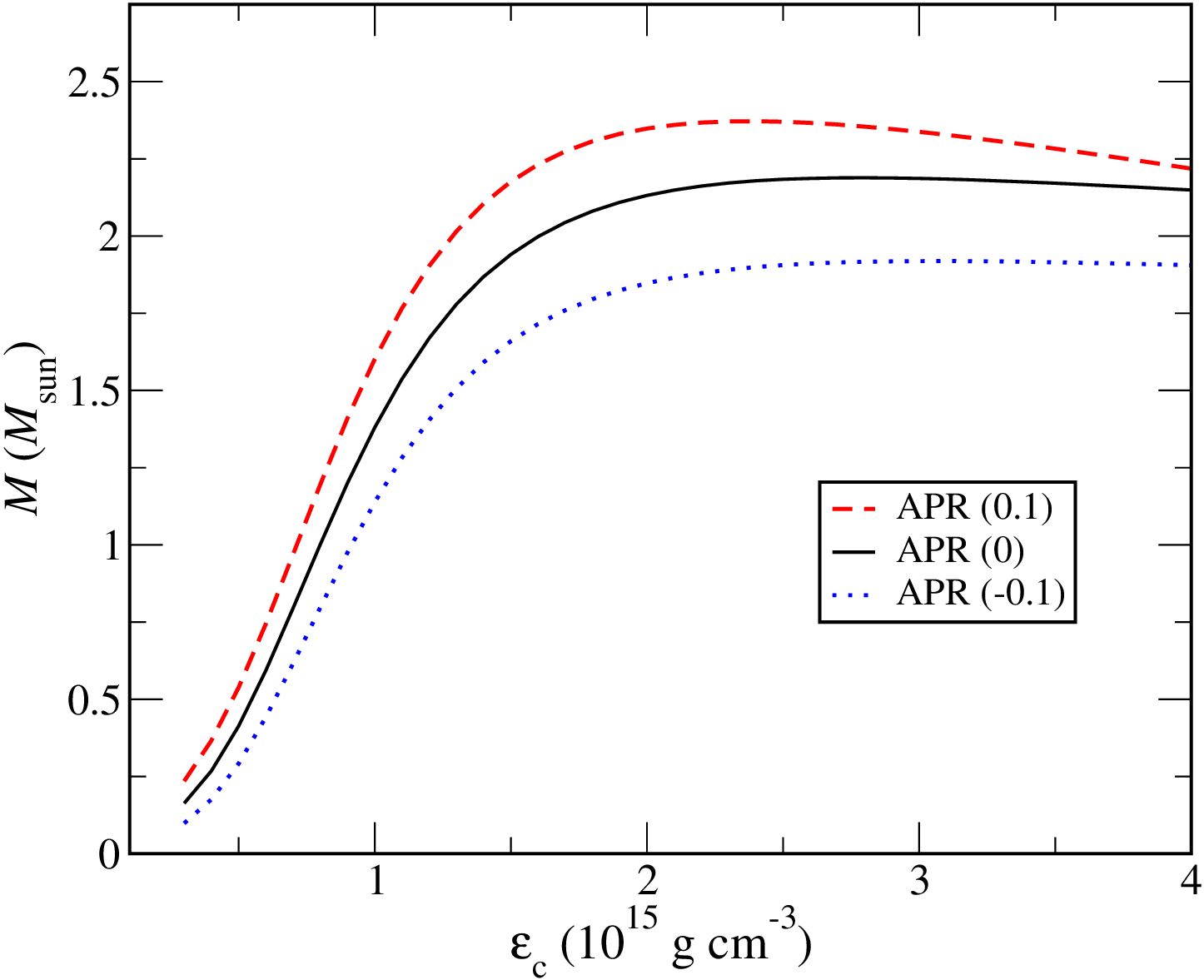}
  \caption{Gravitational mass $M$ plotted against the central density
$\epsilon_c$ for the APR EOS. The values of the coupling parameter 
$8\pi\kappa\epsilon_0$ (with $\epsilon_0=10^{15} {\rm g\ cm}^{-3}$) are 
shown in parentheses. Note that $\kappa = 0$ corresponds to the GR limit.}
  \label{fig:M_rhoc}
\end{figure}

\section{$f$-mode universality} 
\label{sec:fmode}

The oscillation modes of NSs are damped by the emission of 
gravitational waves. They are in general called quasi-normal modes and 
each mode has a complex eigenfrequency $\omega = \omega_{\rm r} + i 
\omega_{\rm i}$. 
The imaginary part $\omega_{\rm i}$ corresponds to the damping rate of the 
oscillation mode. As discussed in Section~\ref{sec:intro}, various attempts 
have been made to find relationships relating $\omega_{\rm r}$ and 
$\omega_{\rm i}$ of the $f$-mode to the global parameters of the star. 
Most of these works \citep{Andersson:1996p20,Andersson:1998p1059,
Benhar1999:p797,Benhar:2004:p124015,Tsui:2005p1029} used $M$ and $R$ as 
the parameters in the analysis. 
However, \citet{Lau:2010p1234} have recently established two 
EOS-independent relations using the parameters $M$ and $I$. The physical 
motivation for replacing $R$ by $I$ in the study of \citet{Lau:2010p1234} 
is that $R$ is sensitive to the low-density part of the EOS, while $I$ 
measures the mass distribution of the star globally. As the dynamics of 
$f$-mode oscillations are signigicantly affected by the mass distribution, 
it is thus expected that $I$ should relate more directly to the $f$-mode.

\begin{figure}
  \centering
  \includegraphics[width=7.5cm]{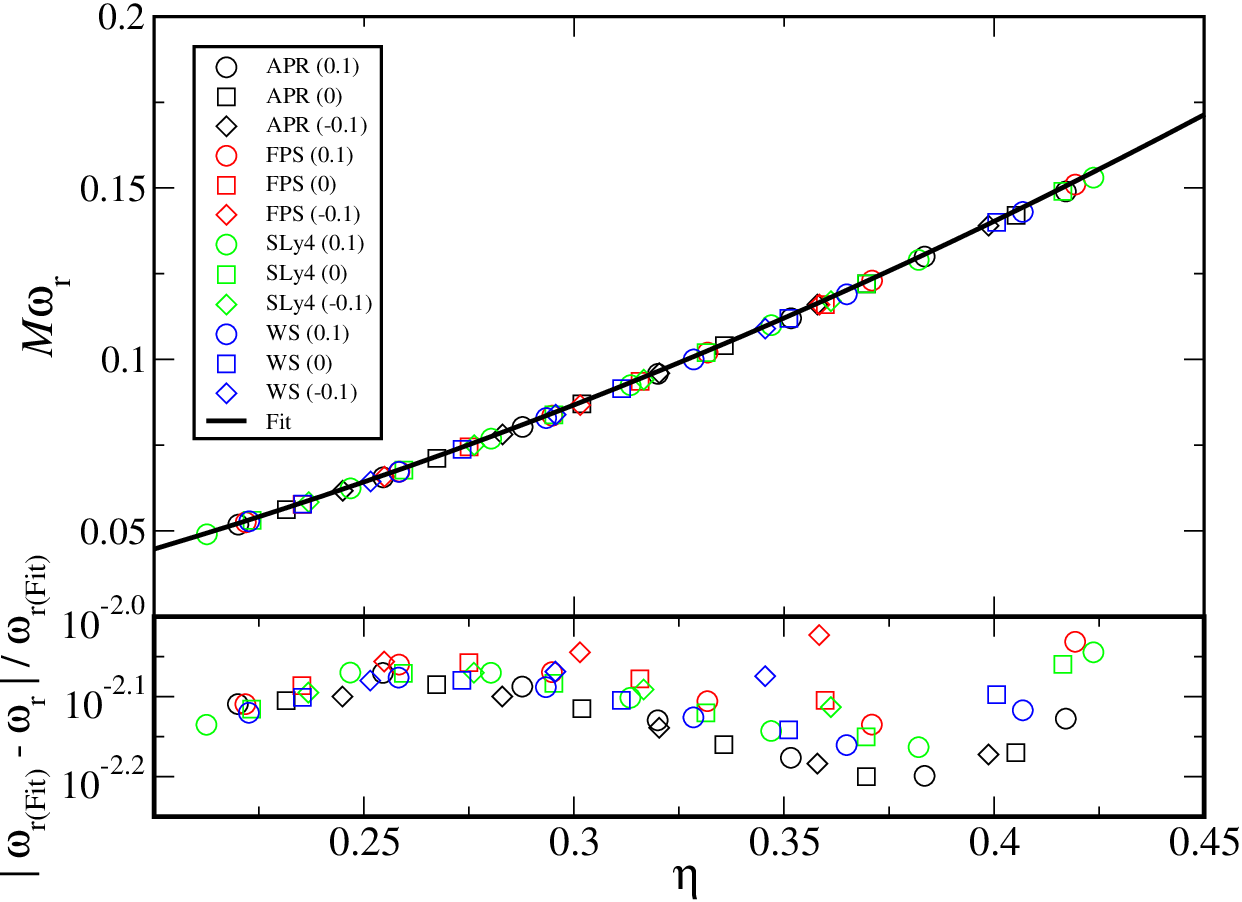}
  \caption{
Real part of the scaled $f$-mode frequency $M\omega_{\rm r}$ for our 
chosen EOS models (APR, FPS, SLy4, and WS) plotted against the effective 
compactness $\eta$ defined in the text. 
Similar to Figure~\ref{fig:M_rhoc}, the values of 
$8\pi\kappa\epsilon_0$ are shown in parentheses. The solid line 
is the fitting curve from Equation~(\ref{eq:fmode_real}). 
The lower panel shows the relative fractional difference between the numerical 
results and the fitting curve. }
  \label{fig:omega_r}
\end{figure}

The universal relations found by \citet{Lau:2010p1234} are given 
by\footnote{Note that we have corrected a typographical error in Equation 
(6) of \citet{Lau:2010p1234}.  }
\begin{equation}
M \omega_{\rm r} = - 0.0047 + 0.133 \eta + 0.575 \eta^2 , 
\label{eq:fmode_real}
\end{equation}
\begin{equation}
I^2 \omega_{\rm i} /M^5 = 0.00694 - 0.0256 \eta^2 , 
\label{eq:fmode_img}
\end{equation}
where the dimensionless factor $\eta \equiv \sqrt{M^3 / I }$. 
Equations~(\ref{eq:fmode_real}) and (\ref{eq:fmode_img}) are much improved 
universal relations in the sense that these relations are less sensitive 
to the EOS, comparing with previous universal relations that use $R$ as a 
parameter. 
We refer the reader to Table 1 in \citet{Lau:2010p1234} for the accuracy of 
Equations~(\ref{eq:fmode_real}) and (\ref{eq:fmode_img}).

Our aim in this section is to study whether the universal 
relations (Equations~(\ref{eq:fmode_real}) and (\ref{eq:fmode_img})) 
found for $f$-mode 
oscillations in GR remain valid in EiBI gravity. 
To this end, we calculate the $f$-mode frequency for NSs in EiBI 
gravity using the apparent-EOS approach as discussed in 
Section~\ref{sec:EiBI_apparentEOS}. 
The procedure is to (1) construct an apparent EOS $\tilde{P}(\tilde{\epsilon})$
for a given coupling parameter $\kappa$ and physical EOS $P(\epsilon)$;
(2) construct an equilibrium background stellar model using the TOV equations
in GR with the apparent EOS; and (3) calculate the $f$-mode frequency for the 
perturbed background model using the numerical codes that we previously 
developed in GR \citep{Lau:2010p1234}, but with $q_{\mu\nu}$ as the 
fundamental metric. 
The method for calculating oscillation modes of NSs in GR
is well established and documented 
\citep[for a review see][]{Kokkotas:1999review}.
We refer the reader in particular to \citet{Lindblom:1983p73} and 
\citet{Detweiler:1985p12} for the formulation we used in the calculations. 

In \citet{Lau:2010p1234}, nine ordinary nuclear-matter and two 
quark-matter EOS models were considered in establishing the universal 
relations (Equations~(\ref{eq:fmode_real}) and (\ref{eq:fmode_img})). 
In this work, 
we consider four nuclear-matter EOS models: model APR \citep{Akmal:98p1804}, 
model FPS \citep{Lorenz:93p379}, model SLy4 \citep{Douchin:2000p107}, 
and model WS \citep{Lorenz:93p379,Wiringa:98p1010}. Among these 
four models, the model APR was also used in \citet{Lau:2010p1234}. 
For the coupling parameter $\kappa$ in EiBI gravity, we consider three 
different values defined by $8\pi\kappa\epsilon_0 = -0.1,\ 0,\ 0.1$, 
where $\epsilon_0 = 10^{15}{\rm g\ cm}^{-3}$. 
These values are consistent with the constraint set by the existence of NSs 
\citep{Avelino:2012p104053}. In particular, $\kappa=0$ corresponds to the GR 
limit. 
In order to show that the range of $\kappa$ we consider is already large 
enough that the resulting NSs in EiBI gravity differ significantly from those 
in GR, we plot the gravitational mass $M$ as a function of the central 
density $\epsilon_c$ for the APR EOS in Figure~\ref{fig:M_rhoc}.
It can be seen from the figure that $M$ can change by as much as $\sim$30\% 
when $8\pi\kappa\epsilon_0$ increases from $-0.1$ to 0.1.

\begin{figure}
  \centering
  \includegraphics[width=7.5cm]{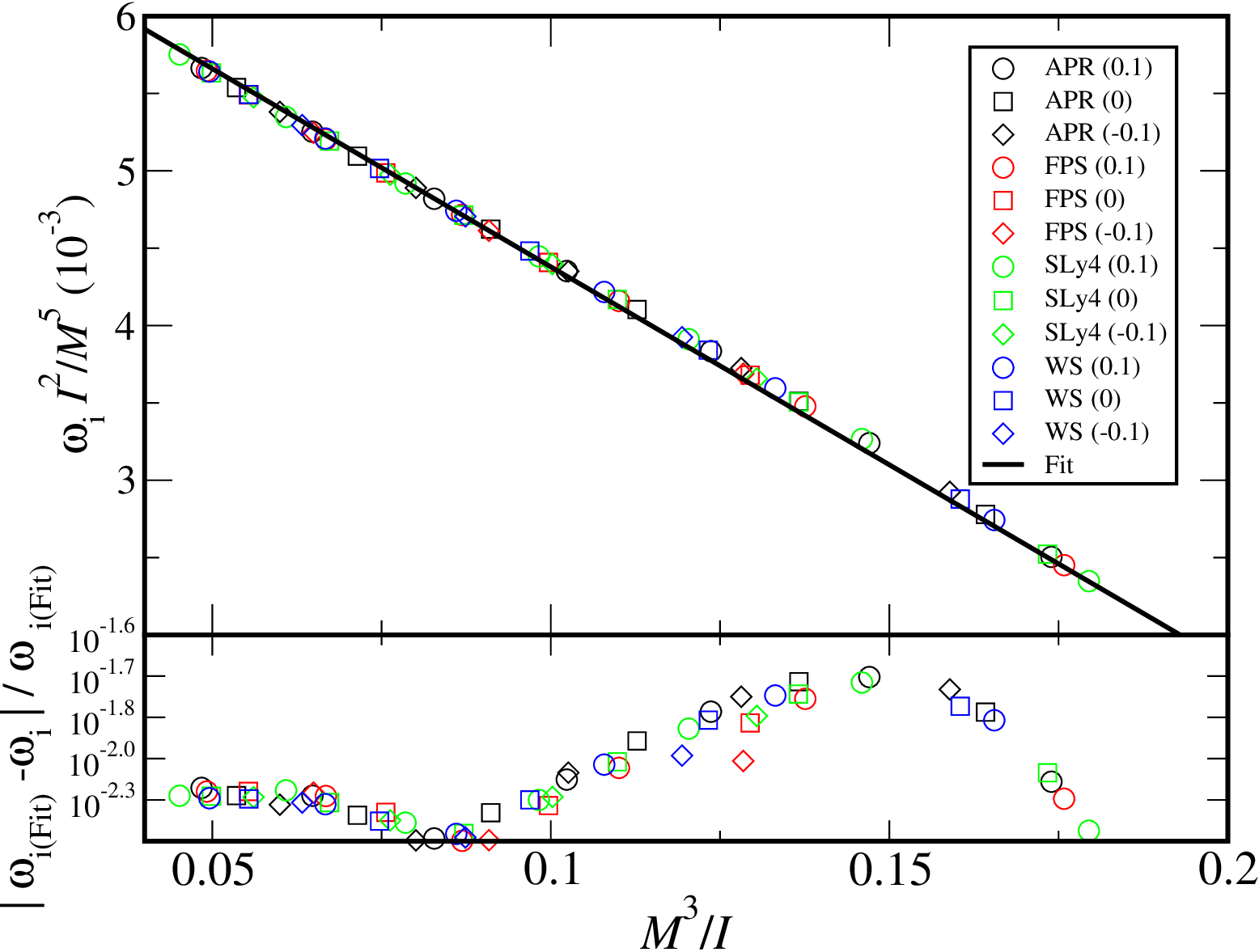}
  \caption{$\omega_{\rm i} I^2 / M^5$ plotted against $\eta^2$
(i.e., $M^3/I$). The solid line is the fitting curve of  
Equation~(\ref{eq:fmode_img}). 
The lower panel shows the relative fractional difference 
between the numerical results and the fitting curve. } 
  \label{fig:omega_i_linear}
\end{figure}

Figure~\ref{fig:omega_r} plots the real part of the scaled $f$-mode frequency
$M\omega_{\rm r}$ against $\eta$ for our chosen EOS models with different 
values of $8\pi\kappa\epsilon_0$.
It can be seen clearly that the data display universal relations that are 
essentially independent of the EOS models and the value of $\kappa$, 
as long as $\kappa$ is within the range constrained by the existence of NSs as 
discussed above. We see that the data can be fit well by 
Equation~(\ref{eq:fmode_real}). The relative fractional difference between the 
numerical results and Equation~(\ref{eq:fmode_real}) is shown in the lower 
panel of the figure. 
Similarly, we plot $\omega_{\rm i} I^2/ M^5$ against $\eta^2$ in
Figure~\ref{fig:omega_i_linear} and see that the data can be modeled 
well by Equation~(\ref{eq:fmode_img}). 
We refer the reader to \citet{Lau:2010p1234} for the motivation of this way 
of plotting $\omega_{\rm i}$. 
In summary, we find that the universal relations 
(Equations~(\ref{eq:fmode_real}) and (\ref{eq:fmode_img})) 
for the $f$-mode of NSs in GR still hold for 
EiBI gravity as long as the coupling parameter $\kappa$ is in the range 
$ | 8\pi\kappa \epsilon_0 | \lesssim 0.1$.

\section{I-Love-Q relations} 
\label{sec:ILoveQ}

After discussing the universality of $f$-mode oscillations, we now turn to 
the universal I-Love-Q relations discovered more recently 
by \citet{Yagi:2013long,Yagi:2013}. 
The moment of inertia of a star is defined by $I \equiv J/\Omega$, 
where $J$ and $\Omega$ are the angular momentum and angular velocity
of the star, respectively. Physically, $I$ determines how fast a star
can spin for a given angular momentum. 
It thus seems quite natural that $I$ should 
somehow relate to the spin-induced quadrupole moment $Q$ of the star, since 
$Q$ characterizes the deformation of the star due to self rotation. 
However, it is surprising that the relation between $I$ and $Q$ found by 
\citet{Yagi:2013long,Yagi:2013} is EOS independent. 
On the other hand, the tidal Love number $\lambda_{\rm tid}$ measures the 
deformation of a NS due to the presence of a companion and is defined by 
$Q_{ij} \equiv - \lambda_{\rm tid} {\cal E}_{ij}$, where $Q_{ij}$ is the 
traceless quadrupole moment tensor of the star and ${\cal E}_{ij}$ is the 
tidal tensor that induces the deformation 
\citep[see, e.g.,][]{Flanagan:08p021502}.
In general, there is no reason why there should exist EOS-independent 
universal relations relating the three quantities $I$, $Q$, and 
$\lambda_{\rm tid}$. 
More specifically, the universal relations concern the dimensionless 
quantities ${\bar I} \equiv I/M^3$, ${\bar Q} \equiv - Q/(M^3 \chi^2)$ 
(with $\chi\equiv J/M^2$ being the dimensionless spin parameter), and 
${\bar \lambda}_{\rm tid} \equiv \lambda_{\rm tid}/ M^5$. 

The relevance of the I-Love-Q relations to astrophysics, 
gravitational-wave, and fundamental physics has been discussed
\citep{Yagi:2013long,Yagi:2013}. For instance, it was proposed that, 
for a detected gravitational wave signal emitted by an inspiralling NS binary, 
the relations could break the degeneracy between the NS quadrupole moment   
and the NS's individual spins. It has also been shown by \cite{Maselli:2013}
that while the I-Love relation connecting ${\bar I}$ and 
${\bar \lambda}_{\rm tid}$ depends on the inspiral 
frequency during the inspiral, it nevertheless remains EOS independent. 
More recently, \citet{Haskell:1309.3885} showed that the I-Q 
universal relation fails for magnetized NSs with long spin periods
($P > 10$ s) and strong magnetic fields ($B > 10^{12}$ G)
\footnote{After we submitted the paper, there were three 
preprints discussing the I-Q universal relation in rapidly rotating 
stars \citep{Doneval:1310.7436,Pappas:1311.5508,Chakrabarti:1311.6509}. }.

As discussed in Section~\ref{sec:intro}, our aim is to study whether 
the universal I-Love-Q relations remain valid in EiBI gravity. 
The methodologies to calculate $I$, $\lambda_{\rm tid}$, and $Q$ in GR 
are well established \citep{Hartle:67p1005,Hartle:68:p907,Flanagan:08p021502,
Hinderer:08p1216,Damour:09p084035,Binnington:09p084018,Yagi:2013long,
Urbanec:13p1903}. 
Here we determine these quantities in EiBI gravity by solving the relevant 
equations in GR with apparent EOSs.

\begin{figure}
  \centering
  \includegraphics[width=7.5cm]{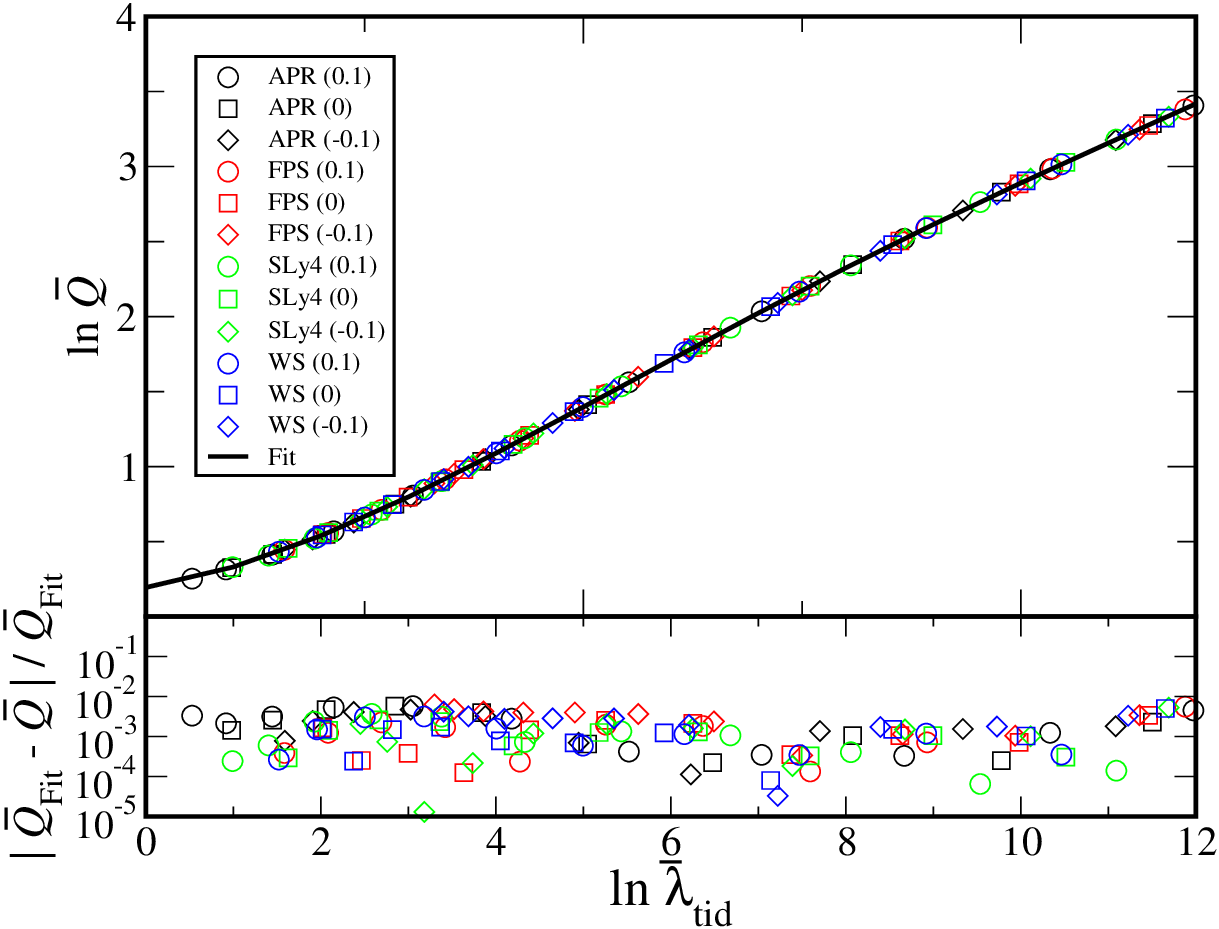}
  \caption{  In the upper panel, $\ln{\bar Q}$ is plotted against 
$\ln{\bar \lambda_{\rm tid}}$ for different EOS models and coupling 
constant $\kappa$. The solid line is the fitting curve 
(Equation~(\ref{eq:ILoveQ_fit})) proposed by \citet{Yagi:2013long,Yagi:2013}.
The lower panel shows the relative fractional difference between the 
numerical results and the fitting curve. } 
\label{fig:Q_Love}
\end{figure}

\begin{figure}
  \centering
  \includegraphics[width=7.5cm]{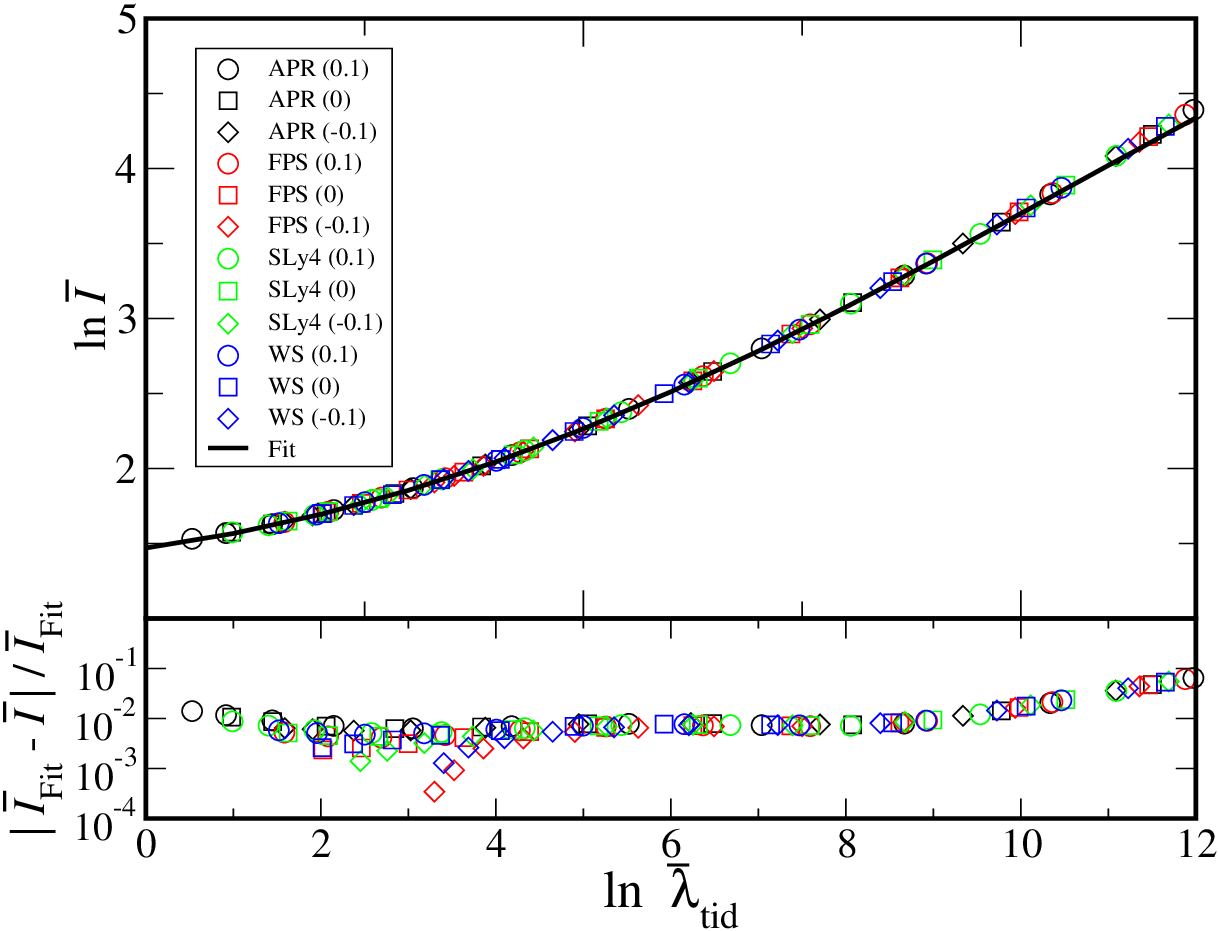}
  \caption{ Similar to Figure~\ref{fig:Q_Love}, but for the plot of 
$\ln{\bar I}$ vs. $\ln{\bar \lambda_{\rm tid}}$.  }
  \label{fig:I_Love}
\end{figure}

\begin{figure}
  \centering
  \includegraphics[width=7.5cm]{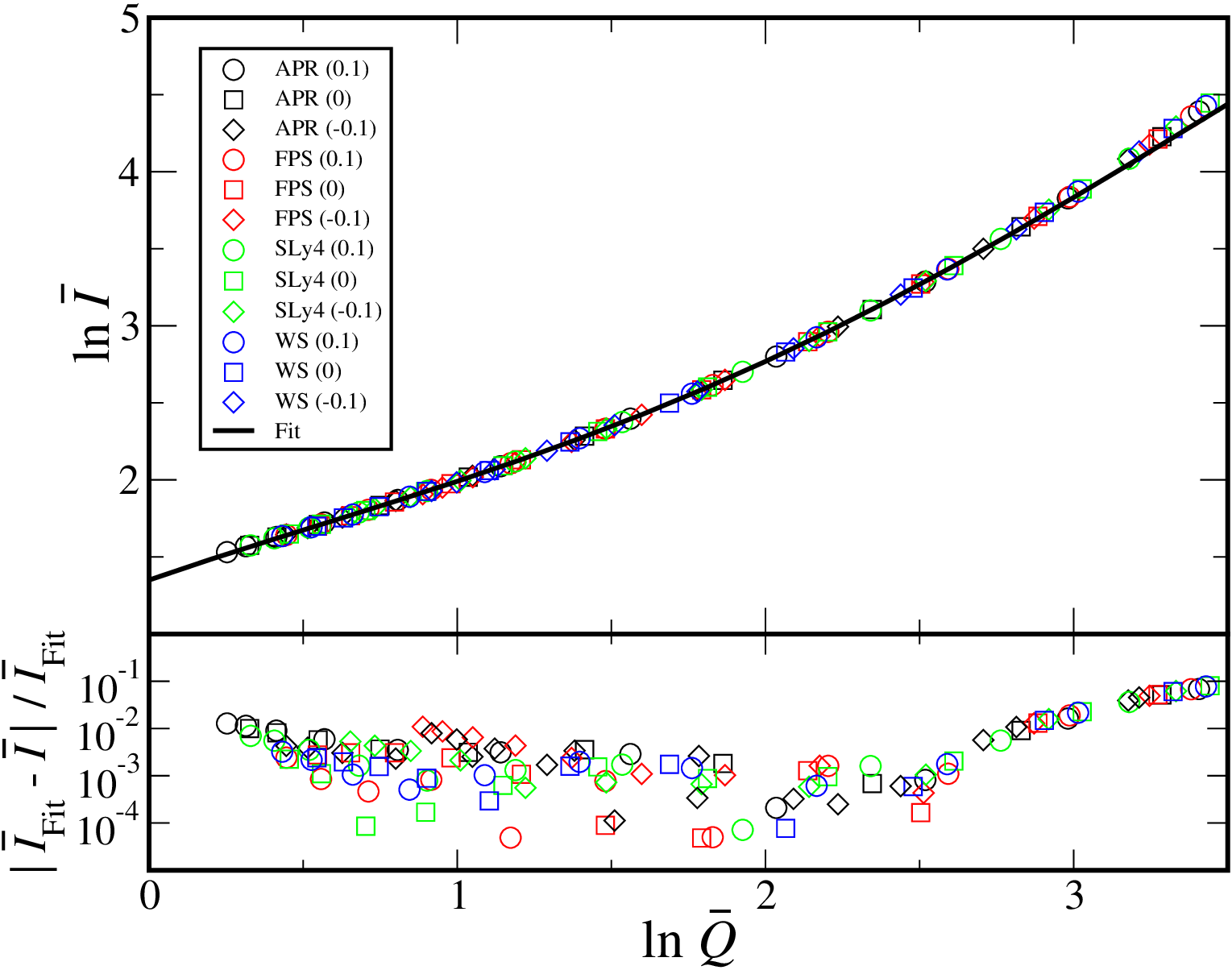}
  \caption{Similar to Figure~\ref{fig:Q_Love}, but for the plot of 
$\ln{\bar I}$ vs. $\ln{\bar Q}$. }
  \label{fig:I_Q}
\end{figure}

In Figures~\ref{fig:Q_Love} and \ref{fig:I_Love}, we plot $\ln{\bar Q}$
and $\ln{\bar I}$ against $\ln {\bar \lambda_{\rm tid}}$, respectively, for
our chosen EOS models and values of $8\pi\kappa \epsilon_0$.  As in 
Section~\ref{sec:fmode}, we consider three different values 
$8\pi\kappa \epsilon_0 = -0.1,\ 0,\ 0.1$. 
Similarly, we plot $\ln{\bar I}$ against $\ln{\bar Q}$ in 
Figure~\ref{fig:I_Q}.  
It is seen clearly from the figures that the I-Love-Q relations 
are essentially independent of the chosen EOS models as discovered by 
\citet{Yagi:2013long,Yagi:2013}. Furthermore, as in the case of the 
$f$-mode universal relations, we find that the I-Love-Q relations 
are insensitive to the coupling parameter $\kappa$ as long as it is in the 
range $|8\pi\kappa\epsilon_0| \lesssim 0.1$. We also note that our numerical 
results can be fit well by the following relation (the solid line in each 
figure) as suggested by \citet{Yagi:2013long,Yagi:2013}: 
\begin{equation}
\ln y_i = a_i + b_i \ln x_i + c_i \left( \ln x_i \right)^2 
+ d_i \left(\ln x_i \right)^3 + e_i \left( \ln x_i \right)^4 , 
\label{eq:ILoveQ_fit}
\end{equation}
where $a_i$, $b_i$, $c_i$, $d_i$, and $e_i$ are some fitting coefficients. 
The relative fractional difference between our numerical results and 
Equation~(\ref{eq:ILoveQ_fit}) is shown in the lower panel of each figure.

\citet{Yagi:2013long,Yagi:2013} also studied the universal relations 
in a modified theory called dynamical Chern-Simons (CS) gravity. 
They found that there also exists a universal I-Love relation 
connecting ${\bar I}$ and ${\bar \lambda_{\rm tid}}$, 
although the relation is different from the corresponding I-Love relation in 
GR. 
They suggested that if one can measure $I$ of the double-binary pulsar
J0737-3039 to 10\% accuracy and $\lambda_{\rm tid}$ to 60\% 
accuracy with future gravitational wave observations, the CS theory can then 
be constrained much better than current tests by six orders of magnitude. 
For comparison, the universal I-Love-Q relations in EiBI theory are the same 
as the GR ones for the range of the coupling parameter $\kappa$, which has 
already been constrained astrophysically. 
Hence, contrary to CS theory, EiBI gravity is an example of a modified
theory where the I-Love-Q relations are degenerate with the corresponding GR 
relations and cannot be used to put a stronger constraint on the theory than 
that obtained by the current astrophysical one.

\section{Conclusions}
\label{sec:conclude}

In this paper, we have studied the EOS-independent universal $f$-mode 
\citep{Lau:2010p1234} and I-Love-Q \citep{Yagi:2013long,Yagi:2013} 
relations for NSs in EiBI gravity. 
With the coupling parameter of the theory in the range 
$|8\pi\kappa\epsilon_0| \lesssim 0.1$, which is constrained by the 
existence of NSs \citep{Avelino:2012p104053}, we find that the universal 
relations discovered in GR remain valid in EiBI gravity.  

Naively, since EiBI gravity reduces to GR when the coupling constant 
$\kappa$ vanishes, one might worry that the agreement between the universal 
relations in EiBI gravity and those in GR is simply because the values of 
$\kappa$ we consider are so small that the effect of 
nonlinear matter-gravity coupling in EiBI gravity is not apparent. 
But this is not the case as we have seen in Figure~\ref{fig:M_rhoc}.  
The effect of nonlinear matter-gravity coupling is not vanishingly small in 
our study. 
On the other hand, since EiBI gravity can be recast to GR with an apparent 
EOS, the nonlinear matter-gravity coupling in EiBI gravity is thus
mimicked by the stiffness of the apparent EOS in GR. 
As long as the stiffness of the resulting apparent EOS is not  
significantly different from realistic nuclear EOS models, which is the 
case in our study, it is then not surprising that the universal relations 
in the two theories are identical. 
For comparison, there also exists a nearly EOS-independent I-Love relation 
in dynamical CS gravity \citep{Yagi:2013long,Yagi:2013}, although the relation 
is different from the GR one.

While we have only focused on the range $|8\pi\kappa\epsilon_0| \lesssim 0.1$
for the coupling parameter in this paper, we have in fact also considered a 
much larger range of $\kappa$ and found that the universal relations are still 
valid. For instance, for $8\pi\kappa\epsilon_0 = 10$, the relative fractional 
differences between the numerical results and the fitted universal curves 
still remain about 0.5\%.
Nevertheless, it should be pointed out that such a large value of $\kappa$ 
would lead to stellar configurations that are very different from typical 
NSs observed in the universe 
(e.g., their masses can be much larger than $1 M_\odot$). 
On the other hand, for a negative value of $\kappa$, we found that stellar 
configurations cannot be constructed if $\kappa$ is too negative since 
$\tilde P$ cannot decrease continuously to zero as one integrates the stellar
equations and hence the boundary conditions at the stellar surface cannot
be fulfilled \citep[see also][]{Sham:2013}.
The exact value of $\kappa$ at which this would happen depends on the EOS, but 
it is typically $8\pi \kappa \epsilon_0 \lesssim - 0.2$ for our chosen EOS 
models.

Finally, let us conclude with two remarks: (1) as discussed in 
Section~\ref{sec:intro}, EiBI gravity is appealing because it reduces to 
GR in vacuum and can avoid some of the singularities that plague GR. 
If it was not due to the fact that EiBI gravity suffers from some 
pathologies associated with compact stars \citep{Pani:2012p251102,Sham:2013}, 
the theory would have one additional merit toward being a serious contender 
to GR because it has essentially the same non-trivial universal 
$f$-mode and I-Love-Q relations in the high-density regime as those in 
GR.  
It leaves one to wonder whether there exists any variation of EiBI gravity or 
more generically a version of nonlinear matter-gravity coupling theory that 
could capture all the nice features of EiBI gravity without suffering from 
any pathologies. 
(2) It should be noted that the reason why there exist universal $f$-mode and 
I-Love-Q relations is not yet understood even in GR. 
\citet{Yagi:2013long,Yagi:2013} suggested two possible reasons for the 
I-Love-Q relations: (1) the physical quantities considered depend most 
strongly on the NS outer layer, where the ignorance of the EOS is smaller than 
that in the NS core and (2) the relations approach the black hole limit as the 
NS compactness increases, and hence the relations do not depend on the 
internal stellar structure. 
It is worth investigating whether these two suggestions are indeed the 
correct explanation for the I-Love-Q relations. It will also be interesting to 
study whether the two apparently different sets of $f$-mode and I-Love-Q 
universal relations have a common origin or not. We hope to return to these 
issues in the future.






\end{document}